\documentclass[12pt]{revtex4-1}
\usepackage{amsfonts,amsmath,amssymb,xcolor}
\usepackage{amsthm,amscd,ulem,bm}
\usepackage{graphicx}
\newcommand{\bib}[2]{\frac{\partial {#1}}{\partial {#2}}}
\newcommand{\bbib}[3]{\frac{\partial^2 {#1}}{\partial {#2}{\partial {#3}}}}
\def\w{{\wedge}}

\begin{document}
\title{
Variational principle of relativistic perfect fluid
}
\author{Takayoshi Ootsuka}
\email{ootsuka@cosmos.phys.ocha.ac.jp}
\affiliation{Physics Department, Ochanomizu University, 2-1-1 Ootsuka Bunkyo, Tokyo, Japan}
\affiliation{NPO Gakujutsu-Kenkyu Network}
\author{Muneyuki Ishida}
\email{ishida@phys.meisei-u.ac.jp}
\affiliation{Department of Physics, Meisei University,
 2-1-1 Hodokubo, Hino, Tokyo 191-8506, Japan }
\author{Erico Tanaka}
\email{erico@sci.kagoshima-u.ac.jp}
\affiliation{Department of Mathematics and Computer Science, 
Kagoshima University, 1-21-35 K\={o}rimoto Kagoshima, Kagoshima, Japan}
\author{Ryoko Yahagi}
\email{yahagi@hep.phys.ocha.ac.jp}
 \affiliation{Physics Department, Ochanomizu University, 2-1-1 Ootsuka Bunkyo, Tokyo, Japan }
\date{\today}

\begin{abstract}

We reformulate the relativistic perfect fluid system on curved space-time. 
Using standard variables, the velocity field $u$,
energy density $\rho$ and pressure $p$, 
the covariant Euler-Lagrange equation is obtained from variational principle.
This leads to the Euler equation 
and the equation of continuity 
in reparametrization invariant form.

\end{abstract}


\maketitle

\section{Introduction}

In cosmology, matter is normally treated as relativistic fluid 
and its equation of motion is studied directly. 
Nevertheless, in cases such as discussing symmetry, 
it is more convenient to utilize Lagrangian formulation, 
and consider the fluid equation via the variational principle.

Naturally, 
various works have been proposed (for example~\cite{Taub}) 
to construct Lagrangian formulation of relativistic 
perfect fluid.
However, while the choice of dynamical variables is essential to simplify the discussion, 
these methods are somehow indirect, such as 
representing the fluid elements by field variables that do not appear
explicitly in the equations of motion. 
Such roundabout may not be desirable for both pedagogical and theoretical reasons, 
{\it e.g.} building a simple cosmological model.  
The generalization of variational principle to the fluid mechanics 
is carried out by following the case of a rigid body~\cite{Marsden-Ratiu}, 
leading to Euler-Poincar\'e equations.
In such sense, the appropriate variables for the case of a perfect fluid are, 
$(u,\rho,p)$, representing: 
velocity vector field of a matter flow, 
energy/mass density, pressure: respectively. 

In fact, 
such way of construction in case of non-relativistic perfect fluid 
is given by Lanczos~\cite{Lanczos}, 
and the case of relativistic dust fluid ($p=0$) is given 
in the elementary text book by Dirac~\cite{Dirac}.
Combining these two approaches, the authors presented 
the Lagrangian formulation of relativistic perfect fluid 
in a geometrical way~\cite{OYIT}, (section 3.3). 
Nevertheless, the above formalism still had some unnecessary complications. 
In the present paper,  
we will present a more simplified and straightforward formulation 
which will give covariant equations of motion. 
This formulation is in accordance with the Eulerian specification of fluid dynamics.

\section{Variational Principle}

The Lagrangian of the perfect fluid on $(n+1)$-dimensional Lorentzian manifold
$(M,g)$ is given by 
\begin{eqnarray}
L_{\rm p.f.}(u,\rho,p) &:=& L_{\rm dust}(u,\rho) + \int\ p\ \delta L (u) \ ,
\label{eq1}
\end{eqnarray}
where the first term is the Lagrangian of dust fluid, $u \in \frak{X}(M)$ is the vector field
of the fluid velocity, 
and $\rho \in \Lambda^{n+1}(M)$ is the mass density $(n+1)$-form of the fluid. 
It is given by 
\begin{eqnarray}
 L_{\rm dust}(u,\rho) &:=& \rho  F(u),
 \label{eq2}
\end{eqnarray}
where we consider 
a Finsler metric of the form: 
$F=\sqrt{g_{\mu\nu}(x)dx^\mu dx^\nu}$, and $F(u)=\sqrt{g_{\mu\nu}(x)u^\mu u^\nu}$ .
Here, $\{(x^0,x^1,x^2,\dots, x^n)\}$ are the local coordinates of $M$
and $\displaystyle u=u^\mu \bib{}{x^\mu}$. 
$F$ and its derivatives are regarded as functions on $TM$, 
and throughout this paper, notation such as $F(u)$ 
means the contraction of the vector field $u$ 
to $dx^\mu$ in the above sense. 
We also set the speed of light $c=1$. 
The second term of Eq.~(\ref{eq1}) represents the effect of pressure 
$p \in C^\infty (M)$, where the integral $\int$ represents symbolically 
that its variation gives $p\ \delta L (u)$. 
$L (u)$ is given by
\begin{eqnarray}
L (u) &:=& \omega_g \,  F(u),
\label{eq3}
\end{eqnarray}
where   
$\omega_g = \sqrt{-g}dx^0\wedge dx^1\wedge \cdots\wedge dx^{n}$,
($g:= {\rm det}\big( g_{\mu\nu}(x) \big)$) is the volume element.

The action is given by the integral 
\begin{eqnarray}
 {\cal A}[u,\rho,p]:=\int_{\sigma} L_{\rm p.f.}(u,\rho,p),
\label{eq4}
\end{eqnarray}
where $\sigma$ represents a $(n+1)$-dimensional submanifold of $M$.
By considering its variation, 
\begin{eqnarray}
 \delta {\cal A}[u,\rho,p]&:=& \int_\sigma \delta L_{\rm p.f.}(u,\rho,p)
 =\int_\sigma \delta L_{\rm dust}(u,\rho) + p\ \delta L(u),
 \label{eq5}
\end{eqnarray}
one obtains the equations of motion. 

\subsection{Equation of dust fluid}
First, we calculate $\delta L_{\rm dust}(u,\rho)$. 
The fluid element shifts infinitesimally to the direction
 $\xi \in \frak{X}(M)$, and 
the corresponding variations of 
$\rho$ and $u$ are represented by the Lie derivatives as,
\begin{eqnarray}
 \delta \rho &=& {\cal L}_\xi \rho, \quad 
 \delta u={\cal L}_\xi u=[\xi, u]=-{\cal L}_u \xi\ , \quad 
 \delta u^\mu=[\xi, u]^{\mu}.
 \label{eq6}
\end{eqnarray}
Then, 
\begin{eqnarray}
 \delta L_{\rm dust}(u,\rho) &=&  (\rho+\delta \rho)F(u+\delta u)
 -\rho F(u) 
  = \delta \rho F(u)+\rho \bib{F}{u^\mu}
  \delta u^\mu
  \nonumber\\ 
  &=&({\cal L}_\xi \rho) F(u)+\rho \bib{F}{u^\mu}
   \langle dx^\mu, {\cal L}_\xi u \rangle
   = (d\iota_\xi \rho ) F(u)-\rho \bib{F}{u^\mu}
    \langle dx^\mu,{\cal L}_u \xi \rangle
   \nonumber\\
   &=&
   d\left\{\iota_\xi \rho F(u)\right\}
   +(-1)^{n+1}
   \iota_\xi \rho \w d(F(u))
   -{\cal L}_u \left(\rho \bib{F}{u^\mu}\xi^\mu\right)
   +\left\langle {\cal L}_u \left(\rho \bib{F}{u^\mu}dx^\mu\right),\xi
   \right\rangle
   \nonumber\\
   &=&
   d\left\{\iota_\xi \rho F(u)-\iota_u \rho \bib{F}{u^\mu}\xi^\mu
    \right\}
   +(-1)^{n+1}
   \iota_\xi \{\rho \w d(F(u))\}
   \nonumber\\
    && -\rho \langle d(F(u)), \xi \rangle
      +\left\langle {\cal L}_u \left(\rho \bib{F}{u^\mu}dx^\mu\right),\xi
   \right\rangle.
 \label{eq7}
 \end{eqnarray}
The Cartan's formula ${\cal L}_u=d\iota_u+\iota_u d$ is used,  
and for notational convenience, we abbreviated 
$\displaystyle \bib{F}{u^\mu}:=\bib{F}{dx^\mu}(u)$. 
The bracket $\langle \, , \rangle$ denotes contraction, for instance, 
\begin{eqnarray*}
  \left\langle {\cal L}_u \left(\rho \bib{F}{u^\mu}dx^\mu\right),\xi
  \right\rangle 
  := \left({\cal L}_u \rho \bib{F}{u^\mu}\right ) \langle dx^\mu, \xi\rangle 
  + \rho \bib{F}{u^\mu} \langle {\cal L}_u dx^\mu ,\xi \rangle,  
\end{eqnarray*}
and $\langle dx^\mu ,\xi \rangle :=\xi^\mu$.
Then since $\rho \w d(F(u))=0$, 
the variational principle tells us that 
\begin{eqnarray}
0&=&{\cal L}_u \left(\rho \bib{F}{u^\mu}dx^\mu\right)-\rho d(F(u)), \nonumber\\
  &=& ({\cal L}_u \rho)\bib{F}{u^\mu}dx^\mu+\rho {\cal L}_u
   \left(\bib{F}{u^\mu}dx^\mu\right)-\rho d(F(u)), \nonumber\\
  &=&  ({\cal L}_u \rho)\bib{F}{u^\mu}dx^\mu
   +\rho \left\{
   \bbib{F}{x^\nu}{u^\mu}u^\nu dx^\mu+\bbib{F}{u^\alpha}{u^\mu}
    u^\nu \bib{u^\alpha}{x^\nu}dx^\mu+\bib{F}{u^\alpha} \bib{u^\alpha}{x^\mu}dx^\mu
   \right\} \nonumber \\
  && -\rho \left(
   \bib{F}{x^\mu}dx^\mu+\bib{F}{u^\alpha}\bib{u^\alpha}{x^\mu}dx^\mu
    \right), \nonumber\\
  &=& ({\cal L}_u \rho)\bib{F}{u^\mu}dx^\mu
   -\rho
   \left[\left\{\bib{F}{x^\mu}-d\left(\bib{F}{dx^\mu}\right)\right\}(u)
    \right] dx^\mu,
    \label{eq8}
\end{eqnarray}
for arbitrary $\xi\in \frak{X}(M)$. 
This equation (\ref{eq8}) is called the Euler-Poincar\'e equation for the dust fluid.
Therefore, clipping off $dx^\mu$, we get,
\begin{eqnarray}
 0&=&\left [({\cal L}_u \rho)\bib{F}{dx^\mu}
   -\rho \left\{\bib{F}{x^\mu}-d\left(\bib{F}{dx^\mu}\right)\right\} \right](u).
\label{eq9}
\end{eqnarray}
Contracting (\ref{eq9}) with $u^\mu$ gives 
$ 0=({\cal L}_u \rho) F(u)$,
where we used the homogeneity 
$u^\mu \bib{F}{dx^\mu}(u)= F(u)$, and its Euler-Lagrange derivative, 
\begin{eqnarray}
 0&=&
   u^\mu \left\{ \bib{F}{x^\mu}-d\left(\bib{F}{dx^\mu}\right)\right\}(u).
\label{eq10}
\end{eqnarray}
Thus, in the case $F(u)\neq 0$, we obtain
\begin{eqnarray}
 {\cal L}_u \rho=0,  && \quad
  \rho  \left\{\bib{F}{x^\mu}-d\left(\bib{F}{dx^\mu}\right)\right\}(u)=0,
\label{eq11}
\end{eqnarray}
where the first equation represents the conservation of mass, while 
the second is the Euler equation for dust fluid. 
This equation is generalized to perfect fluid
in the next section.

Up to now, we have only used the homogeneous property of $F$, 
and not its concrete form given by the Lorentizan metric.  
Therefore, we may choose for $F$ any general Finsler metric 
that satisfies ${\rm rank}\left(\bbib{F}{dx^\mu}{dx^\nu}\right)=n$ . 
In that case, the equation (\ref{eq11}) could be regarded as 
an equation of dust fluid on Finsler spacetime.

\subsection{Equation of perfect fluid}

Here we will calculate 
the pressure term $p\, \delta L(u)$ in Eq.~(\ref{eq5}).
\begin{eqnarray}
 p\, \delta L(u)
 &=& p\, \delta(\omega_g F(u)) =
  p\,  \left\{ \omega_g \left(  F(u+\delta u) - F(u) \right) +\delta\omega_g \, F(u) \right\},
\label{eq12}
\end{eqnarray}
where the first term is calculated similarly as in (\ref{eq7}), 
\begin{eqnarray}
 && p \omega_g \left(  F(u+\delta u) - F(u) \right) \nonumber\\
 &=& p \omega_g \frac{\partial F}{\partial u^\mu} \delta u^\mu 
   = p \omega_g \frac{\partial F}{\partial u^\mu} \langle dx^\mu, \delta u\rangle
   = p \omega_g \frac{\partial F}{\partial u^\mu} \langle dx^\mu,{\cal L}_\xi u\rangle
   =-p \omega_g \frac{\partial F}{\partial u^\mu} \langle dx^\mu,{\cal L}_u \xi\rangle
 \nonumber
 \\
 &=&
 -{\cal L}_u\left( p \omega_g \bib{F}{u^\mu} \langle dx^\mu,\xi \rangle \right)
 + \left \langle {\cal L}_u \left( p\omega_g \bib{F}{u^\mu} dx^\mu \right), \xi \right \rangle
 \nonumber \\
&=& -d \left\{
 \iota_u \left( p\omega_g  \frac{\partial F}{\partial u^\mu} \xi^\mu \right)
 \right\}
 + \xi^\mu \left\{ 
 {\cal L}_u \left( p\omega_g \right) \frac{\partial F}{\partial u^\mu}
 +  p\omega_g\  \iota_u d \left( \frac{\partial F}{\partial u^\mu}\right)
 +p\omega_g \frac{\partial F}{\partial u^\rho} 
 \bib{u^\rho}{x^\mu}
 \right\},
\label{eq13}
\end{eqnarray}
and the second term of Eq.~(\ref{eq12}) becomes
\begin{eqnarray}
 pF(u)\delta\omega_g &=& {\cal L}_\xi\left( pF(u)\omega_g \right)
   - {\cal L}_\xi\left( pF(u) \right) \omega_g \nonumber\\
  &=& d \left\{ \iota_\xi \left( pF(u) \omega_g  \right) \right\}
   - \xi^\mu \left\{ \frac{\partial p}{\partial x^\mu} \omega_g F(u) 
   + p \omega_g \left( \frac{\partial F}{\partial x^\mu}(u)
   + \frac{\partial F}{\partial u^\rho} \frac{\partial u^\rho}{\partial x^\mu} \right)
  \right\}.
  \label{eq14}
\end{eqnarray}
Then,
\begin{eqnarray}
 p \, \delta L(u)
 &=& d \left\{ \iota_\xi\left( p\omega_gF(u) \right)
      -\iota_u \left( p\omega_g \frac{\partial F}{\partial u^\mu} \xi^\mu \right) \right\}
      \nonumber\\
 && + \xi^\mu \left[ 
     {\cal L}_u(p\omega_g) \bib{F}{u^\mu}
      -p\omega_g \left\{ \frac{\partial F}{\partial x^\mu}(u)
     -\iota_u d\left(\frac{\partial F}{\partial u^\mu}\right) \right\}
     -\frac{\partial p}{\partial x^\mu}\omega_g F(u)\right].
 \label{eq15}
\end{eqnarray}
Finally, by using 
Eqs.~(\ref{eq5}), (\ref{eq7}), and (\ref{eq15}),
we obtain
\begin{eqnarray}
 \delta L_{\rm p.f.}(u,\rho,p)
 &=& \delta L_{\rm dust}(u,\rho)
 + p\, \delta L(u) \nonumber
 \\
 &=& d\left\{ \iota_\xi\rho F(u) -\iota_u \rho\ \frac{\partial F}{\partial u^\mu}\xi^\mu
  +p\iota_\xi \omega_gF(u) 
  -p\iota_u \omega_g \frac{\partial F}{\partial u^\mu}\xi^\mu  \right\} 
  + \xi^\mu {\cal EL}_\mu (L_{\rm p.f.}).\ \ \ \ 
 \label{eq16}
\end{eqnarray}
By the variational principle, 
the last term gives the Euler-Lagrange equation, 
\begin{eqnarray}
 {\cal EL}_\mu (L_{\rm p.f.})&=&0, \nonumber \\ 
 {\cal EL}_\mu (L_{\rm p.f.}) &:=& \frac{\partial F}{\partial u^\mu}{\cal L}_u(\rho+p\omega_g)
 -(\rho+p\omega_g) \left\{ \frac{\partial F}{\partial x^\mu} 
 - d \left(\frac{\partial F}{\partial dx^\mu}\right)
 \right\}(u)
 -\frac{\partial p}{\partial x^\mu} \omega_g F(u).\ \ \ \ \ 
 \label{eq17}
\end{eqnarray}
The above equation (\ref{eq17}) is invariant with respect to the transformation 
$(u,\rho,\omega_g) \to (\phi u, \phi^{-1}\rho,\phi^{-1} \omega_g)$, 
where $\phi$ is an arbitrary positive definite function on $M$. 
This transformation $u\to \phi u$ corresponds to changing ``time''
which is a parameter of fluid elements.
Therefore, this scale transformation means reparametrization
in the perspective of Eulerian specification of flow.

\subsection{covariant Euler equation}

In the previous sections, we have scarcely used the concrete form of the 
function $F$, 
which was given by the 
Lorentzian metric. 
In fact, only the homogeneous property of $F$ 
was required to carry out the above calculations. 

Here, we will show that the covariant
Euler-Lagrange (Euler-Poincar\'e) equation of perfect fluid
(\ref{eq17})
will indeed give a familiar Euler form (evolution equation form).

Considering the concrete form,
\begin{eqnarray}
 F(u) &:=& \sqrt{g_{\mu\nu}(x)u^\mu u^\nu}, \quad
 \frac{\partial F}{\partial u^\mu} = \frac{g_{\mu\nu} u^\nu}{F(u)},\ \ \ 
 \rho =:\mu \omega_g, 
 \label{eq21}
\end{eqnarray}
the factor in the second term of 
Eq.~(\ref{eq17}) is rewritten as
\begin{eqnarray}
 &&\left[ \bib{F}{x^\mu}-d\left(\bib{F}{dx^\mu}\right) \right](u)
 =
 \left[ \frac{1}{2F} \bib{g_{\alpha\beta}}{x^\mu}dx^\alpha dx^\beta
 -d\left(\frac{g_{\mu\beta}dx^\beta}{F}\right) \right](u)
 \nonumber\\
 &=&  \frac{1}{2F(u)} \bib{g_{\alpha\beta}}{x^\mu}u^\alpha u^\beta
 -\frac{g_{\mu\beta}du^\beta}{F(u)}
 -\frac{1}{F(u)}\bib{g_{\mu\beta}}{x^\alpha}u^\alpha u^\beta
 +\frac{g_{\mu\beta}u^\beta}{F(u)^3}
 \left(\frac12\bib{g_{\alpha\nu}}{x^\rho}u^\alpha u^\nu u^\rho
 +g_{\alpha\nu}u^\alpha du^\nu \right)
 \nonumber\\
 &=&-\frac{g_{\mu\beta}}{F(u)}
 \left\{ u^\nu \bib{u^\beta}{x^\nu} + {\varGamma^\beta}_{\nu \rho} u^\nu u^\rho \right\}
 +\frac{g_{\mu\beta}u^\beta}{F(u)^3}
 \left\{ u_\nu \bib{u^\nu}{x^\alpha} u^\alpha 
 + u_\nu {\varGamma^\nu}_{\alpha \rho} u^\alpha u^\rho \right\}
  \nonumber\\
 &=&-\frac{g_{\mu\beta}}{F(u)}\nabla_u u^\beta
  +\frac{g_{\mu\beta}u^\beta g_{\alpha\nu}u^\alpha}{F(u)^3} \nabla_u u^\nu
 =
 -\bbib{F}{dx^\mu}{dx^\beta}(u) 
 \nabla_u u^\beta.
 \label{eq22}
\end{eqnarray}
Then, Eq.~(\ref{eq17}) becomes, 
\begin{eqnarray}
 (\rho+p\omega_g)F_{\mu\nu}(u)
 (\nabla_u u^\nu)=\bib{p}{x^\mu}\omega_g F(u) - 
 \bib{F}{u^\mu}
 {\cal L}_u (\rho+p\omega_g),
 \label{eq23}
\end{eqnarray}
where we set
\begin{eqnarray}
 F_{\mu\nu} &:=& \frac{\partial^2 F}{\partial dx^\mu\ \partial dx^\nu}, 
 \quad F_{\mu\nu}(u)=\bbib{F}{dx^\mu}{dx^\nu}(u).
 \label{eq24}
\end{eqnarray}

We would like to solve this equation for $\nabla_u u^\nu$. 
Since $F_{\mu\nu}$ satisfies the homogeneity relation $F_{\mu\nu}dx^\nu =0$ 
or $F_{\mu\nu}(u) u^\nu=0$, it is not invertible. 
Nevertheless, contracting Eq.~(\ref{eq23}) by $u^\mu$, we get
\begin{eqnarray}
 0 &=& {\cal L}_u (\rho+p \omega_g) - ({\cal L}_u p) \omega_g,
\label{eq25}
\end{eqnarray}
where we have assumed $F(u)\neq 0$. 
This is the equation of continuity for the perfect fluid on a Lorentzian manifold, 
and satisfying this equation guarantees that 
a particular solution for (\ref{eq23}) exists. 
Using (\ref{eq25}), the equation (\ref{eq23}) is rewritten as
\begin{eqnarray}
 F_{\mu\nu}(u)
 (\nabla_u u^\nu) &=& \frac{1}{(\mu + p)}\left(
 F(u) \bib{p}{x^\mu}- 
 \bib{F}{u^\mu}
 u(p) \right).
 \label{eq26}
\end{eqnarray}
Define, 
\begin{eqnarray}
  \tilde{F}^{\mu\nu}
 := Fg^{\mu\nu}-\frac{dx^\mu dx^\nu}{F}, 
 \label{eq27}
\end{eqnarray}
which satisfies the relation:
\begin{eqnarray}
 \tilde{F}^{\mu\rho}F_{\rho\nu}=\delta^\mu_\nu-\bib{F}{dx^\nu}\frac{dx^\mu}{F}, \ \ \ 
 \tilde{F}^{\mu\rho}(u)F_{\rho\nu}(u)=\delta^\mu_\nu-
 \bib{F}{u^\nu}
 \frac{u^\mu}{F(u)}.
 \label{eq28}
\end{eqnarray}
Now set, $c_\nu:= {\it r.h.s}$ of (\ref{eq26}), then $\tilde{F}^{\mu\nu}c_\nu$ is the 
particular solution of (\ref{eq23}). Explicitly, it is 
\begin{eqnarray}
 \frac{\tilde{F}^{\mu\nu}(u)}{\mu+p} \left[
 F(u)\bib{p}{x^\nu}-
 \bib{F}{u^\nu}
 u(p) \right]
 =\frac{1}{\mu+p} \left(
 F(u)^2
 g^{\mu\nu}\bib{p}{x^\nu}-u^\mu u(p) \right).
 \label{eq29}
\end{eqnarray}

The general solution would be in the form $\tilde{F}^{\mu\nu}c_\nu+\lambda u^\mu$,  
where
$\lambda=\lambda(x,dx)(u)=\lambda (x^\mu,u^\mu)$ 
is an arbitrary and 
homogeneous function of degree one in the $u^\mu$ variables. 
The general solution includes one free function $\lambda (x,u)$, since
${\rm rank}(F_{\mu\nu})=n=(n+1)-1$, and $(F_{\mu\nu})$ has 0-eigen vector $u^\mu$. 
Thus it is obtained as, 
\begin{eqnarray}
 \nabla_u u^\mu=\frac{1}{\mu+p}\left\{
 F(u)^2
 g^{\mu \nu}\bib{p}{x^\nu}-u^\mu u(p)\right\}+\lambda(x,u)u^\mu,
\label{eq30}
\end{eqnarray}
or
\begin{eqnarray}
 \nabla_u u^\mu=\frac{F(u)^2}{\mu+p} g^{\mu\nu}
 \bib{p}{x^\nu}+\lambda_1(x,u)u^\mu.
\label{eq31}
\end{eqnarray}
The
$\lambda_1(x,u)=\lambda_1(x^\mu,dx^\mu)(u)$ 
also is an arbitrary and  
homogeneous function of degree one in 
$u^\mu$.
The existence of these arbitrary functions $\lambda$, $\lambda_1$ 
indicates that these equations are reparametrization invariant with respect to $u$.
Similarly as in the previous section, the equations (\ref{eq25}) and (\ref{eq31}) 
are unchanged by the transformation $\phi$, namely
$(u,\omega_g) \to 
 (\phi u,\phi^{-1}\omega_g)$.
Therefore, we will call this {\it the covariant
Euler equation of curved spacetime}.

\section{Discussion}

We have presented the Lagrangian formulation of 
the perfect fluid on Lorentzian manifold by using 
standard and minimum variables.
The Euler-Lagrange equation (\ref{eq17}),
is derived geometrically from the Lagrangian (\ref{eq1}) 
defined by a specific Finsler metric given in terms of Lorentizan metric $g$.
We showed that this equation gives the Euler form (\ref{eq31})
and the equation of continuity (\ref{eq25}). 
The above results are invariant 
with respect to the transformation:  
$(u,\rho,\omega_g) \to (\phi u,\phi^{-1} \rho,\phi^{-1} \omega_g)$, 
where $\phi$ is an arbitrary positive definite function on $M$. 
This invariance corresponds to the reparametrization 
invariance of fluid elements with respect to the time parameter.
This allows us more flexibility to choose the variables that may be advantageous 
in considering solutions or symmetries of the fluid field or energy density 
on curved spacetime. 
The function $\lambda (x,dx)$ in Eq.(\ref{eq30}) (or $\lambda_1(x,dx)$ in Eq.~(\ref{eq31}))
can be determined consistently with the choice of the 
evolution rate.
For example, by taking 
the velocity speed,
$F(u)=1,\ u_\nu u^\nu=1$, and 
$u_\nu \nabla_u u^\nu =0$, we obtain, 
$\displaystyle \lambda_1 (x,u)=-\frac{u(p)}{\mu+p}$. 
Then we get
\begin{eqnarray}
 \nabla_u u^\mu= \frac{g^{\mu\nu}-u^\mu u^\nu}{\mu+p} \bib{p}{x^\nu},
 \label{eq33}
\end{eqnarray}
while the continuity equation becomes
\begin{eqnarray}
 \nabla_\alpha \{(\mu+p)u^\alpha\}-u^\mu \bib{p}{x^\mu}=0,
 \label{eq32}
\end{eqnarray}
which are indeed the relativistic Euler equation and 
continuity equation of perfect fluid~\cite{Choquet}.

The discussion in the previous subsections, especially in 
subsection A. and B., 
was established mostly without the reference to the concrete form of 
the metric $F$. The important property which contributed to the theory 
was only the homogeneity of $F$.
This suggests that we may choose for $F$ a more general 
form of Finsler metric 
instead of the specific form given by the Lorentzian metric.
We are now working on such extensions, hoping to report the results soon.

\begin{acknowledgements}

We thank Prof. M. Morikawa and the astrophysics and cosmology lab (Ochanomizu university)
for support and encouragements.
E. Tanaka thanks YITP of Kyoto University.   

\end{acknowledgements}



\end{document}